# Phase transition in the $5d^1$ double perovskite $Ba_2CaReO_6$ induced by high magnetic field

Hajime Ishikawa, Daigorou Hirai, Akihiko Ikeda, Masaki Gen, Takeshi Yajima,
Akira Matsuo, Yasuhiro H. Matsuda, Zenji Hiroi, Koichi Kindo
Institute for Solid State Physics, the University of Tokyo,
5-1-5 Kashiwanoha, Kashiwa, Chiba, Japan
E-mail: hishikawa@issp.u-tokyo.ac.jp

Magnetic properties of an antiferromagnetic double perovskite oxide $Ba_2CaReO_6$, where $Re^{6+}$ ($5d^1$) ions with large spin-orbit coupling are arranged on the face-centered-cubic lattice, are investigated using pulsed high magnetic field up to 66 T. Magnetization and magnetostriction measurements have revealed a magnetic field induced phase transition at around 50 T. The phase transition accompanies a jump of magnetization and longitudinal magnetostriction of approximately $2 \times 10^{-4}$ with the change of power law behavior, indicating sizable coupling between the electronic degrees of freedom and the lattice. The high field phase exhibits a magnetic moment approximately 0.2 $\mu_B$, which is close to the values observed in $5d^1$ double perovskite oxides with non-collinear magnetic structure. We argue that $Ba_2CaReO_6$ is an antiferromagnet that sits close to the phase boundary between the collinear and non-collinear phases, providing the target material for investigating the interplay between spin-orbital entangled electrons and magnetic field.

## I. INTRODUCTION

Physical properties emerging from the combination effect of electron correlation and spin-orbit coupling have attracted considerable attention in the last decade [1]. In $5d$ transition metal ions with partially filled $t_{2g}$ orbital, large spin-orbit coupling of 0.4 eV may entangle the effective orbital angular momentum and the spin angular momentum [1]. The electronic state is described by the effective total angular momentum $J_{eff}$. For example, an unpaired electron in the $d^1$ and $d^5$ ions in the cubic octahedral crystal field occupies the $J_{eff} = 3/2$ quartet and the $J_{eff} = 1/2$ doublet, respectively [1-3]. In solids, interactions among spin-orbital entangled electrons may give rise to exotic ground states such as a spin liquid [2] and multipolar orders [3].

When the $d^1$ ions with the $J_{eff} = 3/2$ state interact on the face-centered-cubic (FCC) lattice, unconventional ground states such as multipolar orders [3], spin-orbit dimer phase [4], and spin-orbital liquid [5] may emerge. In the multipolar ordered states, magnetic dipolar, electric quadrupolar, and magnetic octupolar moments can be the order parameter [3]. Magnetic properties of rhenium double perovskite oxides, where $Re^{6+}$ ($5d^1$) ions are arranged on the FCC lattice, have been investigated as Mott insulators with the $J_{eff}$ = 3/2 state [3,6-8]. For example, cubic $Ba_2MgReO_6$ exhibits an electric quadrupolar order accompanying tiny tetragonal lattice distortions and a magnetic order with a net magnetization [6,7], while tetragonal $Sr_2MgReO_6$ exhibits an antiferromagnetic order without a net magnetization [8]. These examples demonstrate that various multipolar orders can be realized in rhenium double perovskite oxides by applying a certain perturbation such as chemical pressure via non-magnetic cations.

Another way for exploring a novel electronic phase is applying strong magnetic field comparable to the magnitude of interactions among electrons. In the spin-1/2 Heisenberg model on the tetragonally distorted FCC lattice as seen in the rhenium double perovskite oxides at low-temperature, successive field induced magnetic phases including the 1/2-magnetization plateau may appear in the magnetization process [9]. While the electric quadrupoles do not couple with magnetic field directly, a possibility of magnetic field induced quadrupolar transition has been discussed in $CeB_6$ [10], where the Ce atom with $4f^1$ electron configuration has $\Gamma_8$ quartet ground state of which symmetry is identical to the $J_{eff}$ = 3/2 quartet.

Here, we focus on a rhenium double perovskite oxide $Ba_2CaReO_6$, of which fundamental physical properties are reported previously [11]. We performed detailed characterization of the bulk physical properties on the polycrystalline sample by measuring magnetization, heat capacity, and linear thermal expansion, which may detect the lattice distortions in response to the electric quadrupolar or orbital order that represents the changes of the charge distribution. We further performed magnetization and magnetostriction measurements in strong pulsed magnetic field and discovered a novel phase transition at around 50 T. We argue that $Ba_2CaReO_6$ is an antiferromagnet that exhibits an interplay between the spin-orbital entangled electrons and the magnetic field.

## II. EXPERIMENTAL METHODS

Polycrystalline samples of $Ba_2CaReO_6$ and its non-magnetic analogue $Ba_2CaWO_6$ were prepared by a solid-state method. A stoichiometric mixture of BaO, MgO, and $ReO_3$ ($WO_3$) powders was mixed and pelletized in an argon-filled glove box. The pellet was heated at 750 °C for 48 h in an evacuated quartz glass tube. The sintered pellet was crushed and re-pelletized in a glove box, and then heated at 900 °C for 48 h in an evacuated quartz glass tube. The phase purity of the sample was confirmed by the powder x-ray diffraction (XRD) measurements using a diffractometer with Cu K$\alpha$ radiation (SmartLab, RIGAKU). Magnetization measurements between 1.8 and 350 K were performed up to 7 T using a commercial SQUID magnetometer (MPMS3, Quantum Design). Magnetization measurements were performed up to around 66 T by the induction method using a pickup coil in the pulsed magnetic field with the pulse length of 4 millisecond generated at the International MegaGauss Science Laboratory at the Institute for Solid State Physics, the University of Tokyo. The magnetization process of the relative compounds $Ba_2MgReO_6$ and $Sr_2MgReO_6$, which were prepared in the course of the previous studies [6-8], were measured up to around 51 T and compared to that of $Ba_2CaReO_6$. The linear thermal expansion was measured by the strain gauge method. A strain gauge (KFL05-120-C1-11, gauge factor 2.01 at 297 K, Kyowa, Japan) was glued to a sintered powder pellet. The sample size was recorded as a change in the resistance of the strain gauge, which was calibrated by measuring a copper plate (purity 99.99%) with a known linear thermal expansion [12]. The magnetostriction was measured in the pulsed magnetic field with the pulse length of 36 millisecond up to around 60 T by the fiber Bragg grating technique using the optical filter method [13]. The composite sample made of the polycrystalline powder and the epoxy resin (STYCAST 1266) was glued to the optical fiber and the change of the sample size were measured in the pulsed magnetic field. Heat capacity measurements were performed by the relaxation method in the commercial apparatus (PPMS, Quantum Design).

## III. EXPERIMENTAL RESULTS

The powder XRD pattern of $Ba_2MgReO_6$ obtained at 300 K is consistent with the cubic double perovskite structure ($Fm$-$3m$) with $a$ = 8.37446(4) Å (see supplemental material [14]). The lattice constant is very close to $a$ = 8.371(4) Å reported in the previous neutron diffraction study [11]. A structural phase transition at $T_s$ = 130 K into the tetragonal $I4/m$ structure is reported. The powder XRD pattern of our sample at 20 K is consistent with the $I4/m$ structure with $a$ = 5.9018(1) Å and $c$ = 8.3827(1) Å, which are similar to $a$ = 5.896(1) Å and $c$ = 8.375(1) Å reported previously at 17 K. Rietveld refinement of our powder XRD data to determine the O atom positions did not converge to a reliable crystal structure likely due to the small x-ray scattering cross section of O compared to other heavy elements. However, a simulated powder pattern using the atomic positions determined by the previous neutron diffraction measurements exhibits good agreement to our powder XRD pattern [14], suggesting that our sample has essentially identical crystal structure to the previous report.

The temperature dependence of magnetic susceptibility $M/H$ of $Ba_2CaReO_6$ measured at 7 T (Fig.1) is consistent with the previous report [11]. The inverse susceptibility $H/M$ above 150 K behaves linearly as expected from the Curie-Weiss law. The slope of $H/M$ changes at around 130 K, where the cubic-tetragonal transition occurs. The effective magnetic moment $\mu_{eff}$ and Weiss temperature $\Theta$ estimated from the linear fit of $H/M$ between 150 and 350 K is 0.700(1) $\mu_B$ and -41.5(1) K, which are similar to the 0.744(2) $\mu_B$ and -38.8(6) K reported previously. The same fit between 50 and 120 K in the tetragonal phase yields $\mu_{eff}$ and $\Theta$ of 0.659(1) and -23.2(4) K, respectively. The changes of $\mu_{eff}$ and $\Theta$ suggests the changes in the electronic state and the exchange interactions. The $\mu_{eff}$ is substantially smaller than the spin only value for the $d^1$ ion (1.73 $\mu_B$). It is close to the values observed in $5d^1$ rhenium double perovskite oxides such as 0.68 $\mu_B$ in $Ba_2MgReO_6$ [6] and 0.8 $\mu_B$ in $Sr_2MgReO_6$ [8]. $M/H$ exhibits a sharp drop at $T_m$ =16 K indicating an antiferromagnetic order. Stronger drop of $M/H$ at $T_m$ than observed in the previous report may suggest the good sample quality.

The temperature dependence of linear thermal expansion $\Delta L/L_{297K}$ exhibits a clear kink at $T_s$ reflecting the structural changes (Fig.2). $\Delta L/L_{297K}$ also exhibits a kink at $T_m$, indicating the magnetic transition accompanies a lattice distortion. The temperature dependence of the specific heat $C_p/T$ exhibits a hump at $T_s$ (Fig.3(a)), which is emphasized when the data is compared with the non-magnetic analogue $Ba_2CaWO_6$. At $T_m$, $C_p/T$ exhibits a sharp peak with an additional peak at 13 K. The two peaks are observed in another sample batch (#2 in Fig. 3(b)), suggesting the second peak is intrinsic to the compound. $C_p/T$ measured at 0 and 5 T almost completely overlap with each other, indicating the transitions are insensitive to the weak magnetic field. Two-step magnetic orders are seen in some double perovskite oxides with heavy magnetic transition metals such as $Sr_2YRuO_6$ [15], $La_2ZnIrO_6$ [16], and $La_2LiMoO_6$ [17], and the possibilities of the development of short-range spin corelations or rearrangement of the spins are discussed. Note, however, the absence of the anomaly in $M/H$ at 13 K in $Ba_2CaReO_6$ in spite of the presence of a sharp peak in $C_p/T$ is unusual. There might be a certain competing order, while its origin is unclear at present.

The electronic entropy $S_e(T)$ was estimated from the specific heat data. Specific heat of the non-magnetic and isostructural $Ba_2CaWO_6$ was used to estimate the lattice specific heat in the tetragonal phase. $S_e(T)$ estimated by directly subtracting the $C_p/T$ of $Ba_2CaWO_6$ from that of $Ba_2CaReO_6$ is shown by the solid line in the inset of Fig.3(a). $S_e(T)$ amounts to 5.3 J/mol K at $T_m$, which is close to $R\ln2$ = 5.76 J/mol K expected for the Kramers doublet. $S_e(T)$ amounts to 13 J/mol K at $T_s$, which is larger than $R\ln4$ = 11.5 J/mol K expected for the $J_{eff}$ = 3/2 quartet. This is likely due to the underestimation of the lattice contribution. Alternatively, the lattice contribution below 30 K can be estimated by rescaling the $C_p/T$ of $Ba_2CaWO_6$ so as to coincide with that of $Ba_2CaReO_6$ at 30 K (Fig.3(b)). $S_e(T)$ estimated in this way is plotted in the inset of Fig.3(a). The two estimations yield nearly identical result because the lattice specific heat is much smaller compared to the electronic contribution below $T_m$. The agreement of the two results confirms that electronic entropy released at $T_m$ is close to $R\ln2$.

The magnetization of $Ba_2CaReO_6$ was measured in the pulsed high magnetic field by changing the maximum magnetic field values up to 66 T (Fig.4(a)). The magnetization curve at 4.2 K deviates from the linear behavior around 35 T and exhibits a steep increase at around 45 T, indicating the presence of a magnetic field induced phase transition. The slope of the curve appears to become smaller as the maximum magnetic field is increased, evoking the magnetization saturation or a plateau. The full magnetic moment of $Ba_2CaReO_6$ estimated from the $\mu_{eff}$ in the tetragonal phase (0.66 $\mu_B$) and the spin-only value for the $d^1$ ion (1.73 $\mu_B$) is 0.66/1.73 = 0.38 $\mu_B$. The magnetization of 0.2 $\mu_B$ at the high field region is clearly smaller than the full moment.

The magnetization curve exhibits a hysteresis in the increasing and decreasing magnetic fields above 35 T. The hysteresis loop does not close in the magnetic field ranges well above the field induced phase transition up to 66 T. The hysteresis was observed in the magnetostriction measurements shown below, which was performed in the magnetic field with approximately 10 times longer pulse length. Observation of a similar hysteresis in the measurements with different pulse lengths indicates the hysteresis is the intrinsic property. As the temperature is increased, the anomaly and the hysteresis become smaller (Fig.4(b,c)). The transition survives at 14 K below $T_m$ but disappears at 18 K above $T_m$. $M/H$ at 56 T plotted along with the data at 7 T indicates that the transition temperature coincides with $T_m$ (inset of Fig.4(b)).

To obtain an insight on the magnetic field induced phase transition of $Ba_2CaReO_6$, the magnetization process of related $5d^1$ double perovskites $Ba_2MgReO_6$ and $Sr_2MgReO_6$ were examined (Fig.4(a)). Magnetization curve of $Ba_2MgReO_6$ exhibits a steep increase with hysteresis at low magnetic field region as seen in a ferromagnet. The size of the net magnetic moment is approximately 0.2 $\mu_B$, which is smaller than the full moment but much larger than those in typical canted antiferromagnet. This behavior is attributed to the two-sublattice spin structure with large spin canting angle of approximately 40° [7]. The magnetization gradually increases from 0.2 $\mu_B$ at 0.5 T to 0.3 $\mu_B$ at around 50 T. The magnetization curve exhibits a persistent hysteresis up to around 30 T as observed in $Ba_2CaReO_6$ and does not fully saturate even at 50 T. The magnetization of $Sr_2MgReO_6$, which exhibits a collinear antiferromagnetic order without net magnetic moment at 55 K, increases linearly up to 51 T.

The magnetic field induced phase transition in $Ba_2CaReO_6$ was clearly detected in the magnetostriction $\Delta L/L_{0T}$ measured at 4.2 K. We obtained data for two different epoxy composite samples made of the polycrystalline powder from the same sample batch (Data A and B in Fig.5). Two measurements yield qualitatively identical results. The longitudinal magnetostriction is positive and rapidly increases at around 40 T. $\Delta L/L_{0T}$ amounts to $2.1 \times 10^{-4}$ at 59 T (Data A) and $2.9 \times 10^{-4}$ at 53 T (Data B). The transverse magnetostriction, which is measured by bending the optical fiber, is negative and amounts to $-3 \times 10^{-5}$ at 59 T (Data A) and $-1.1 \times 10^{-4}$ at 53 T (Data B). The differences in the magnitude of magnetostriction between the two samples should be attributed to the difference of the density of the polycrystalline powder in the epoxy composite and/or the degree of adhesion between the sample and the optical fiber. The longitudinal magnetostriction is plotted against magnetization in the inset of Fig.5. The magnetostriction changes quadratically to magnetization up to around 40 T. The deviation from the quadratic behavior is observed as increasing the field. $\varDelta L/L_{0T}$ is roughly proportional to $M^{0.5}$ in the high field region.

## IV. DISCUSSIONS

In the compounds containing $d^1$ ion with strong spin-orbit coupling, the $J_{eff} = 3/2$ quartet can be formed as a result of the coupling of the orbital and the spin angular momentum. The effective magnetic moment of the $J_{eff} = 3/2$ quartet is ideally zero, however, finite magnetic moment is induced by the hybridization effect between the $d$ and ligand $p$ orbitals [3]. $\mu_{eff}$ of the $5d^1$ double perovskite materials ranges 0.2-0.8 $\mu_B$ depending on the type of the ligand anion [6-8,18-20]. $\mu_{eff}$ of 0.7 $\mu_B$ in $Ba_2CaReO_6$ is a typical value for the oxide with the $5d^1$ ion and suggests that $J_{eff} = 3/2$ quartet is formed in the high temperature cubic phase.

One expects the release of electronic entropy of $R\ln4$ when the electronic degrees of freedom of the $J_{eff} = 3/2$ quartet are ordered at low temperature. The electronic entropy released at $T_m$ is close to $R\ln2$ expected for the Kramers doublet. This indicates that the degeneracy of the $J_{eff} = 3/2$ quartet is lifted at temperatures above $T_m$. A factor that may lift the degeneracy of the $J_{eff} = 3/2$ quartet is the cubic-tetragonal transition at $T_s$. The tetragonal compression of the ligand octahedra below $T_s$ should stabilize a Kramers doublet with dominant $d_{xy}$ orbital [21]. The temperature dependence of the magnetic susceptibility of the $d^1$ double perovskite in the similar situation is discussed in Ref. [22]. A calculation shows that the size of the Curie-Weiss moment and the magnitude of Weiss temperature becomes smaller when the degeneracy of the $J_{eff} = 3/2$ quartet is lifted by an orbital order. The changes of the Curie-Weiss moment and the Weiss temperature at $T_s$ in $Ba_2CaReO_6$ is consistent with the calculation and support the idea that the degeneracy of the $J_{eff} = 3/2$ quartet is lifted at $T_s$.

According to the theory on $d^1$ double perovskite oxides with strong spin-orbit coupling [3], a non-collinear two-sublattice magnetic order with a net magnetization due to the spin canting, which is called the canted antiferromagnetic order hereafter, or a collinear antiferromagnetic order without a net magnetization may appear depending on the relative magnitude among ferromagnetic and antiferromagnetic superexchange interactions and the electric quadrupolar interaction. The canted antiferromagnetic order is stabilized when sizable ferromagnetic superexchange and electric quadrupolar interactions are present. The ferromagnetic superexchange interaction occurs via orthogonal $d$ orbitals such as $d_{xy}$ and $d_{yz}$ ($d_{zx}$) orbitals, while the antiferromagnetic superexchange interaction occurs via the same type of $d$ orbital [3]. A uniform tetragonal distortion of the meta-ligand octahedra would increase the occupancy of a certain orbital and reduce the superexchange path for the ferromagnetic interaction. The distortions of the $ReO_6$ octahedra estimated from the Re-O bond lengths is 1.5% in $Ba_2CaReO_6$

[11], which is much larger than 0.1% in $Ba_2MgReO_6$ with the canted antiferromagnetic order. The large tetragonal distortion may reduce the ferromagnetic interaction and stabilize the collinear antiferromagnetic order in $Ba_2CaReO_6$. The $ReO_6$ octahedra in $Sr_2MgReO_6$ is approximately 1% distorted. The collinear antiferromagnetic order might be stabilized in $Sr_2MgReO_6$ by the similar reason. The antiferromagnetic order in the $d^1$ double perovskite oxides is expected to accompany the changes of the electric quadrupolar moments [3]. Observation of the kink at $T_m$ in $\Delta L/L_{297K}$ is compatible with the scenario.

Then we discuss the magnetic field induced phase transition in $Ba_2CaReO_6$ at around 50 T. In the conventional spin-flop transition caused by the magnetic anisotropy, the magnetization curve exhibits larger slope at high field region than in the low field region or a direct saturation to the full moment. In the case of the former, the linear extrapolate of the high field magnetization curve should intersect the origin of the $M$-$H$ curve. The slope of the high field magnetization curve of $Ba_2CaReO_6$ estimated at 60-65 T (dashed line in the Fig.3) is $2.0 \times 10^{-3}$ and $8.8 \times 10^{-4}$ $\mu_B$/T in the increasing and decreasing magnetic fields, respectively. While the former is larger than the slope in the low field region ($9.6 \times 10^{-4}$ $\mu_B$/T), the latter is smaller. The linear fits apparently do not intersect the origin of the $M$-$H$ curve. The magnetization of 0.2 $\mu_B$ at 66 T is smaller than the full moment of 0.38 $\mu_B$ estimated from the $\mu_{eff}$. Therefore, the transition is not compatible with the conventional spin-flop transition.

In the spin-1/2 Heisenberg model on the tetragonally distorted FCC lattice, a magnetic field induced 1/2-magnetization plateau may appear [9]. The magnetization of 0.2 $\mu_B$ at 66 T is close to the 1/2 of the full moment estimated from $\mu_{eff}$. Magnetization curve in the plateau state may not be completely flat due to the van Vleck paramagnetic susceptibility. The slope of the magnetization curve estimated at 60-65 T corresponds to $1.1 \times 10^{-3}$ and $4.9 \times 10^{-4}$ emu/mol in the increasing and decreasing magnetic fields, respectively. This is much larger than the van Vleck paramagnetic susceptibility of approximately $2 \times 10^{-4}$ emu/mol reported in the $5d^1$ double perovskite materials [18-20]. Such large slope of the magnetization curve is not compatible with the 1/2-magnetization plateau.

We argue the magnetic field induced canted antiferromagnetic order in $Ba_2CaReO_6$. The size of the magnetic moment of 0.2 $\mu_B$ at high fields is close to the values observed in $5d^1$ double perovskite oxides with canted antiferromagnetic order [6,7,18,23]. The gradual increase of the magnetization and the persistent hysteresis at high fields are similar to the features observed in $Ba_2MgReO_6$ with the canted antiferromagnetic order (Fig.3(a)). In the case of $Ba_2MgReO_6$, the features in the magnetization curve are explained by considering the magnetic responses from magnetic domains formed within the tetragonal domains formed at the quadrupolar order [6]. Similarly complex domains can be formed in $Ba_2CaReO_6$ that exhibits successive cubic-tetragonal transition and a magnetic order. The observation of the magnetic field induced phase transition suggests that $Ba_2CaReO_6$ resides at the parameter regime close to the boundary between the collinear and canted antiferromagnetic orders. On the other hand, the absence of the transition up to 50 T in $Sr_2MgReO_6$ suggests that the compound resides deep inside the collinear antiferromagnetic region. The antiferromagnetic superexchange interaction may be enhanced in $Sr_2MgReO_6$ more effectively compared to $Ba_2CaReO_6$ by the differences in the lattice constants and oxygen positions, which modifies the orbital overlapping relevant to the superexchange interactions.

The size of the longitudinal magnetostriction $\Delta L/L_{0T} \sim 2 \times 10^{-4}$ observed at the magnetic field induced phase transition in $Ba_2CaReO_6$ is larger than the values reported in the single crystals of various spin-1/2 oxide magnets at high magnetic fields [24-26]. Considering the measurements are performed on the powder diluted by the epoxy resin, the actual magnetostriction should be larger. The relatively large magnetostriction in spite of the suppressed magnetic moment due to the formation of the $J_{eff} = 3/2$ state indicates the sizable coupling between the electronic degrees of freedom and the lattice in $Ba_2CaReO_6$. The different sign of the longitudinal and transverse magnetostriction in the powder sample indicates the highly anisotropic nature of the magnetostriction. The ordered orbitals in the tetragonal crystal structure should make the coupling between spin and the lattice anisotropic.

$\Delta L/L_{0T}$ often exhibits a power law behavior $\Delta L/L_{0T} \propto M^p$ with different values of $p$ depending on the material [24-26]. In the conventional exchange striction mechanism that considers the strain dependent exchange interactions and the elastic energy [25,26], magnetostriction is proportional to the expectation value of the local spin correlation $\langle S_i \cdot S_j \rangle$, where $S_i$ and $S_j$ represent the interacting spins in the crystal. The $M^2$ behavior as observed in $Ba_2CaReO_6$ at low fields is compatible with the classical antiferromagnetic picture. We expect the $M^2$ behavior from spins in the canted antiferromagnetic order because it is a kind of two sublattice antiferromagnetic order as in the low field phase. The change of the power law behavior from $M^2$ to $M^{0.5}$ indicates that different mechanism works at high fields. According to the theory [3], different quadrupolar (orbital) orders may appear depending on the magnetic ground state. In the high field phase, the changes in the orbital degrees of freedom may contribute to the relatively large magnetostriction. Further experimental and theoretical studies are necessary for the understanding of the orbital states and the microscopic origin of the $\Delta L/L_{0T} \propto M^{0.5}$ behavior.

## V. CONCLUSION AND PERSPECTIVES

We have investigated the magnetic properties of a $5d^1$ rhenium double perovskite $Ba_2CaReO_6$ and discovered a magnetic field induced transition at around 50 T. The linear thermal expansion and magnetostriction measurements detected the lattice distortions likely related to the changes in the orbital degrees of freedom. The magnetic field induced phase transition is attributed to the transition from the collinear to the canted antiferromagnetic order. $Ba_2CaReO_6$ would reside in the parameter regime close to the boundary between the two magnetic orders, representing a suitable material for investigating the interplay between the spin-orbital entangled electrons and magnetic field. Performing magnetostriction measurements on a single crystal would give further information on the orbital states at high fields. Uncovering the origin of the unusual specific heat peak at 13 K by the microscopic probe such as nuclear magnetic resonance measurements would be another interesting future work.

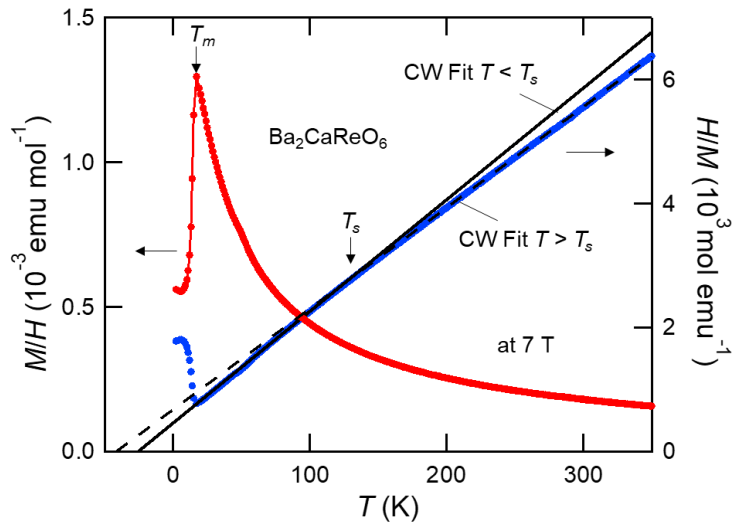

Fig. 1. (a) Temperature dependence of magnetic susceptibility $M/H$ (red) and the inverse susceptibility $H/M$ (blue) of $Ba_2CaReO_6$. The cubic-tetragonal transition at $T_s = 130$ K and the magnetic transition at $T_m = 16$ K are indicated by the arrows. The dashed and solid lines indicate the linear fits at 150-350 K and 50-120 K, respectively.

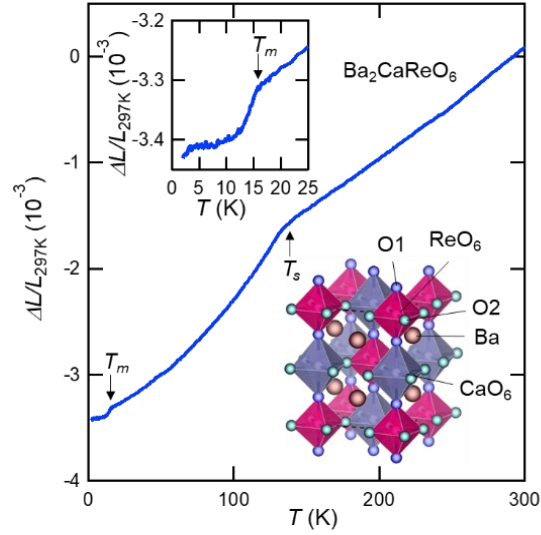

Fig. 2. Linear thermal expansion $\Delta L/L_{297K}$ of $Ba_2CaReO_6$. The data around $T_m$ is shown in the upper inset. The crystal structure below $T_s$ depicted by VESTA [27] is shown in the lower inset. The O atoms at the axial and equatorial position in the $ReO_6$ octahedra (red) are shown by blue and green spheres, respectively.

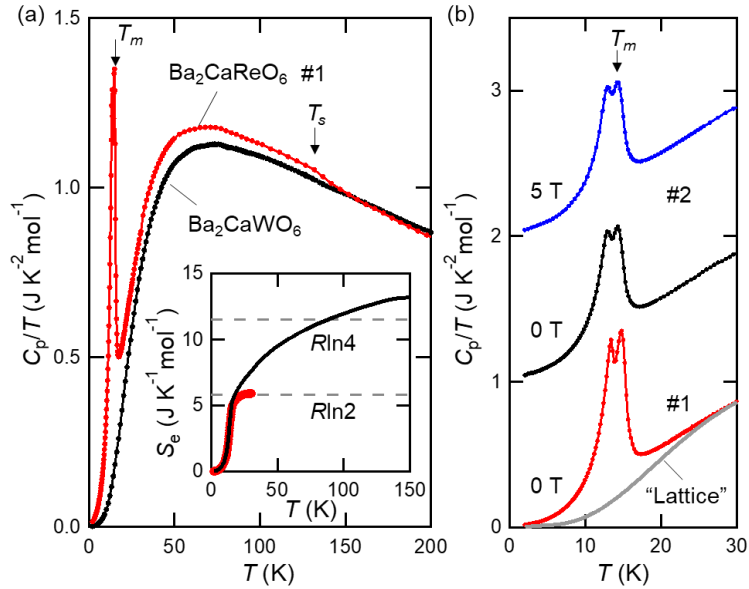

Fig. 3 (a) Temperature dependence of the specific heat divided by temperature $C_p/T$ of $Ba_2CaReO_6$ (red) and its non-magnetic analogue $Ba_2CaWO_6$ (black). The electronic entropy $S_e$ estimated as shown in the main text is shown in the inset. (b) $C_p/T$ of $Ba_2CaReO_6$ below 30 K measured for different samples (#1 and #2). The lattice specific heat estimated by rescaling the data of $Ba_2CaWO_6$ is shown by the gray solid line for the sample #1. $C_p/T$ of the sample #2 is measured at 0 and 5 T: the data is shown with offsets.

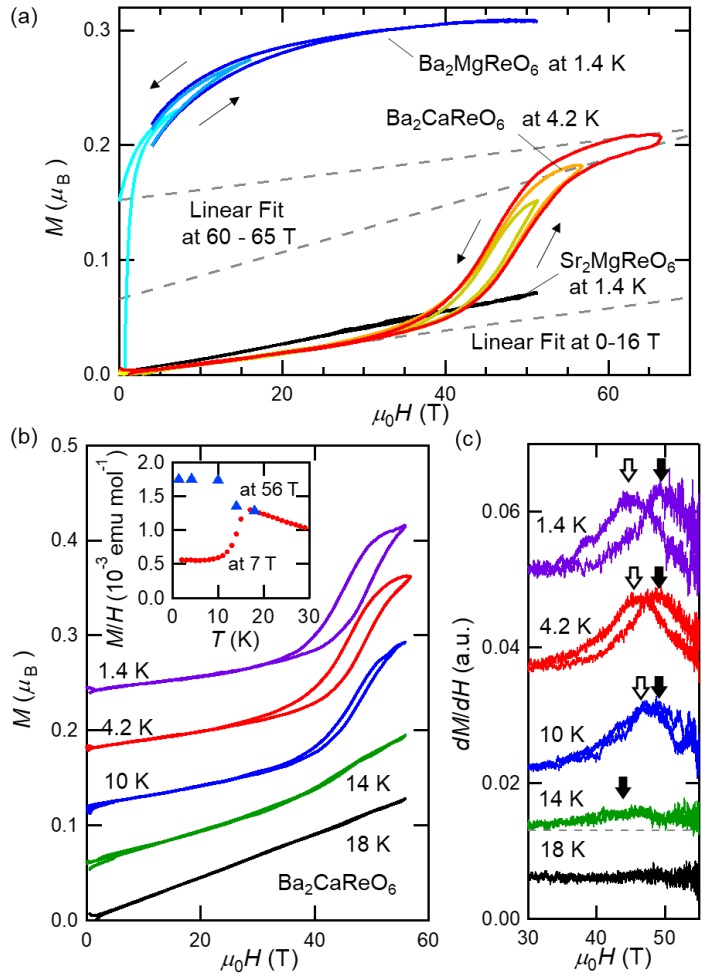

Figure 4. (a) Magnetization curves of $Ba_2CaReO_6$ measured up to 66 T (red) at 4.2 K and those of $Ba_2MgReO_6$ (blue) and $Sr_2MgReO_6$ (black) measured up to 51 T at 1.4 K. The magnetization curve measured with smaller maximum magnet fields are shown for $Ba_2CaReO_6$ and $Ba_2MgReO_6$. The high field data is corrected to so as coincide with the low field SQUID data. The gray dashed line on the magnetization curve of $Ba_2CaReO_6$ is the linear fits at selected magnetic field ranges. (b) Temperature dependence of the magnetization curve of $Ba_2CaReO_6$ measured up to 56 T (left) and (c) its magnetic field derivative $dM/dH$. The data are shown with offsets along the vertical axis. The peaks of $dM/dH$ in increasing and decreasing fields are marked by the filled and unfilled arrows, respectively. The dashed line on the data at 14 K is the guide for the eyes.

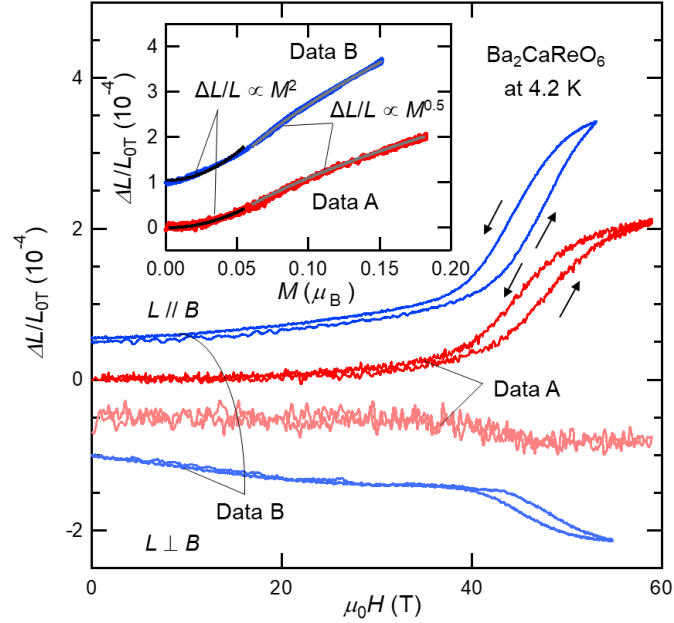

Figure 5 Magnetostriction $\Delta L/L_{0T}$ of $Ba_2CaReO_6$ measured at 4.2 K. Data measured in the increasing and decreasing field processes are shown. Two data sets for different epoxy composite samples are shown. The magnetostriction with positive and negative values correspond to the longitudinal and transverse magnetostriction, respectively. The data are shown with the offsets along the vertical axis. The longitudinal magnetostriction is plotted against the magnetization in the inset. The black and gray lines in the inset is the fit to the power law $\Delta L/L_{0T} \propto M^2$ and $M^{0.5}$, respectively.

Acknowledgement

This work was supported by JSPS Grant-in-Aid Grant Number JP19K23420 and JP20H01858.

# Phase transition in the $5d^1$ double perovskite $Ba_2CaReO_6$ induced by high magnetic field

Hajime Ishikawa, Daigorou Hirai, Akihiko Ikeda, Masaki Gen, Takeshi Yajima,
Akira Matsuo, Yasuhiro H. Matsuda, Zenji Hiroi, Koichi Kindo

Institute for Solid State Physics, the University of Tokyo,
5-1-5 Kashiwanoha, Kashiwa, Chiba, Japan
E-mail: hishikawa@issp.u-tokyo.ac.jp

1. Powder x-ray diffraction pattern of the $Ba_2CaReO_6$ at 300 K and 20 K

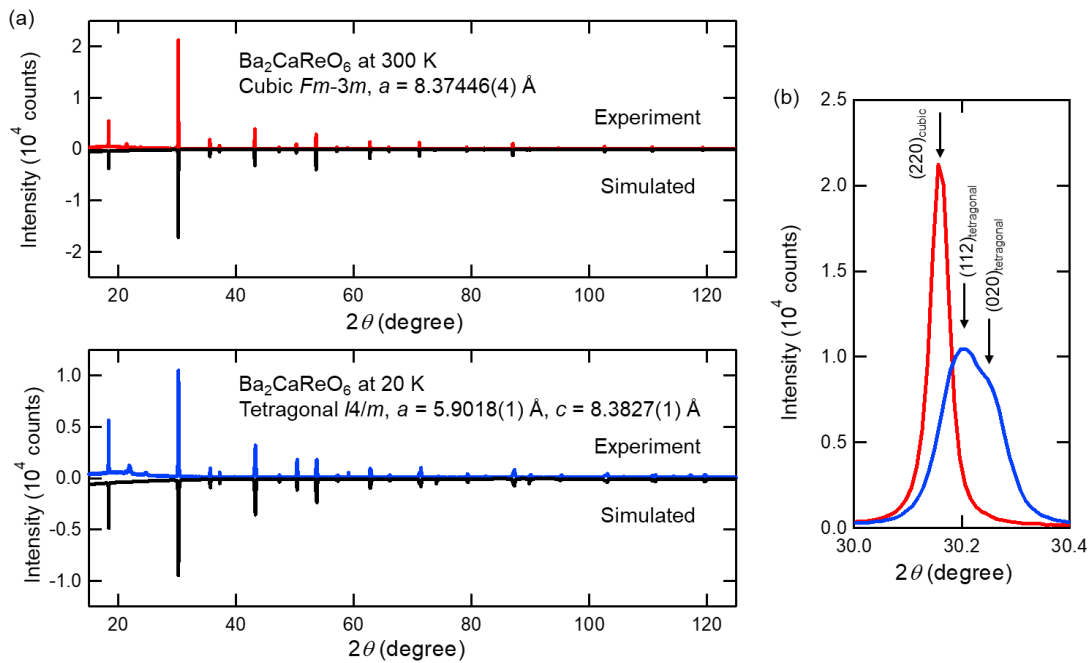

(a) Powder x-ray diffraction pattern of the $Ba_2CaReO_6$ at 300 K and 20 K. The simulated pattern based on the crystal structure determined by the neutron diffraction data [S1] are shown. (b) An example of the peak splitting observed in the powder x-ray diffraction pattern. The (220) reflection in the cubic unit cell splits into the (112) and (020) reflections in the tetragonal unit cell.

[S1] K. Yamamura, M. Wakeshima, and Y. Hinatsu, Structural phase transition and magnetic properties of double perovskites $Ba_2CaMO_6$ ($M$ = W, Re, Os), *J. Solid State Chem.*, 179 605 (2006).

2. Magnetostriction of $Ba_2CaReO_6$ measured at 4.2 and 20 K in pulsed magnetic field

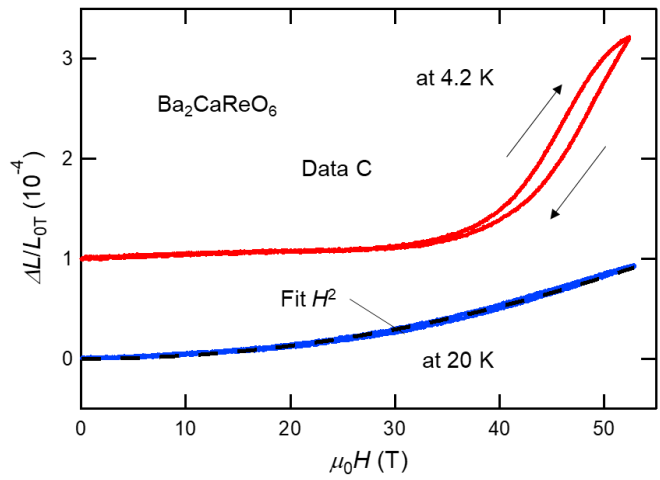

The magnetostriction of $Ba_2CaReO_6$ is measured at 20 and 4.2 K, which are above and below the magnetic transition temperature $T_m$ = 16 K. The sample batch is different from those used in the main text. The anomaly corresponding to the magnetic field induced phase transition is reproduced at 4.2 K. Above $T_m$, the anomaly is not observed and the magnetostriction exhibits a quadratic behavior.